\documentclass[]{aastex}
\usepackage{graphicx}
\usepackage{apjfonts}
\usepackage{amsmath}
\usepackage{natbib}
\usepackage{color}
\usepackage[normalem]{ulem}
\shorttitle{RXTE Study of IGR J19294+1816}
\shortauthors{Roy, Choudhury \& Agrawal}

\begin{document}


\title{Timing and Spectral Study of IGR J19294+1816 with RXTE: Discovery of Cyclotron Feature}


\author{Jayashree Roy\altaffilmark{1,2}, Manojendu Choudhury\altaffilmark{1}, P. C. Agrawal\altaffilmark{1,3}}
\affil{$^{1}$UM-DAE Centre For Excellence In Basic Sciences, Vidyanagari, Santa Cruz (E), Mumbai - 400098}

\email{jayashree.roy@cbs.ac.in}
\altaffiltext{2}{NASI Research Associate}
\altaffiltext{3}{NASI Senior Scientist}


\begin{abstract}
Rossi X-ray Timing Explorer (RXTE)/Proportional Counter Array (PCA) observations of IGR J19294+1816 covering two outburst episodes are reported. The first outburst happened during MJD 54921-54925 (2009 C.E.) and the second one happened during MJD 55499-55507 (2010 C.E.). In both the cases the PCA observations were made during the decay phase of the outburst, with the source exhibiting temporal and spectral evolution with the change in flux. At the bright flux level an absorption feature at 35.5 keV is detected in the spectra which may be attributed to Cyclotron Resonance Scattering Feature corresponding to a magnetic field of $B = 4.13\times10^{12}$ Gauss. This is also detected at a lower significance in two other observations. In addition an Fe line emission at 6.4 keV is prominently detected during the highest flux. X-ray pulsations are detected in 9 out of 10 observations; no pulsations were found in the observation with the lowest flux level. During this observation with the lowest flux the pulsation phenomenon becomes detectable only at the soft X-ray bands. 
\end{abstract}



\section{Introduction}

IGR J19294+1816 was discovered by the INTErnational Gamma-Ray Astrophysics Laboratory (INTEGRAL) using IBIS/ISGRI camera at R.A.=292.42 deg and Dec=+18.28 deg ($\pm$3' at 68\%, J2000) on 2009 March 27 \citep{2009ATel.1997....1T}. Follow up analysis of the Swift archival data by \citet{2009ATel.1998....1R} led them to the conclusion that the source   Swift J1929.8+1821, observed on 2007 December 9 and 13, to be the same as IGR J19294+1816 with an improved position at J2000, RA= 19h 29m 55.9s \& Dec=+18deg 18' 39"($\pm$ 3.5" at 90\%). They detected a periodicity of 12.4 seconds from the power density spectrum showing a feature at $8.04_{-0.05}^{+0.02}\times10^{-2}$ Hz using Swift/XRT data. Their analysis of the timing and the spectral features suggested the source to be an accreting pulsar. \cite{2009ATel.2002....1S} confirmed IGR J19294+1816 to be an accreting pulsar with the presence of 12.44 seconds pulsation from the source using the Rossi X-ray Timing Explorer (RXTE)/Proportional Counter Array (PCA) observation on 2009 March 31. \cite{2009ATel.2008....1C} suggested the source to be a Be/X-ray class of binary system based on its position on the pulse period vs orbital plot of \citet{1986MNRAS.220.1047C}. 
\cite{2009A&A...508..889R} identified an infrared counterpart of the source. From the studies of infrared magnitudes dereddened with different values of the interstellar absorptions, they estimated the source to be at a distance of, d$\gtrsim$8 Kpc. They suggested that the source could possibly be a supergiant fast X-ray transients instead of a Be/X-ray transient because of short ($\sim$ 2000-3000 s) and intense flares that are more typical of supergiant fast X-ray transients. However, during 2010 outburst of the source, the source showed a very smooth and gradual change in flux. Further the spectral parameters during the two months spanned by the Swift observations, along with its low spin period, seem to confirm that the source is a Be/X-ray transient \citep{2011A&A...531A..65B}. 
In this paper, we report the timing and spectral properties of the source during the decay phase of the 2009 and 2010 outburst as observed by the RXTE/PCA. Our spectral analysis reveals the first ever detection of Cyclotron Resonance Scattering Features (CRSF) in the source.

\section{Data Analysis}
RXTE, (PCA) \citep{2006ApJS..163..401J} data of the source were obtained from High Energy Astrophysics Science Archive Research Center (HEASARC) data archive (http://heasarc.gsfc.nasa.gov),  the details of which are presented in Table \ref{tbl-1}. The observed spectral  parameters are presented in Tables \ref{tbl-2} and \ref{tbl-3}. Among the 5 Proportional Counter Units (PCU's) sensitive in 3-60 keV range, only the PCU 2 data (from all the layers) are reported in this paper as it was the only common PCU consistently covering all the observations. 
The X-ray spectra were extracted using standard 2 mode data with 16 second binning and the lightcurves were extracted from the event mode data. The data was filtered with data selection criteria that removed the stretches of observation that had the South Atlantic Anomaly passage time and included the stretches of observations for which the Earth elevation $>$ 10$^\circ$ \& the pointing offset $<$ 0.02$^\circ$. 
Faint background model was used for background estimation of the spectral analysis except for the observation on MJD 55499.46 (Obs.id.: 95438-01-01-00) when the source was just bright enough to warrant the strong background model. The background and dead time corrections were applied to the spectra while barycentric corrections were performed, using the ftools task `fxbary', for all the timing analyses reported here. 
The source lies in the Galactic plane where flux from the Galactic ridge needs to be included in the spectral modeling \citep{1998ApJ...505..134V}. We have strictly followed the recipe of \citet{2009A&A...508..889R} while adding the Galactic ridge spectrum to the instrumental background spectrum. 
The data reduction and analysis was carried out using \texttt{HEASOFT}\footnote{http://heasarc.gsfc.nasa.gov/docs/software/lheasoft/}, which consists of (chiefly) \texttt{FTOOLS} for general data extraction and analysis, \texttt{XRONOS} \citep{1992EMIS..59...59S} for the timing analysis and \texttt{XSPEC} \citep{1996ASPC..101...17A} for the spectral analysis. 

The source was observed for a total of ten PCA pointings, five during the 2009 outburst (MJD 54921.32-54925.83) and five during the 2010 outburst (MJD 55499.46-55507.24). As evident from the flux evolution in Table \ref{tbl-2}, on both occasions the observations are during the decay phase of the outburst. The flux values, measured from the extracted wide band, 3-60 keV, spectra, are comparatively lower during the 2009 outburst (Table \ref{tbl-2}). Details of timing and spectral analysis are discussed in the following subsections.

\subsection{Timing properties}

\begin{figure}
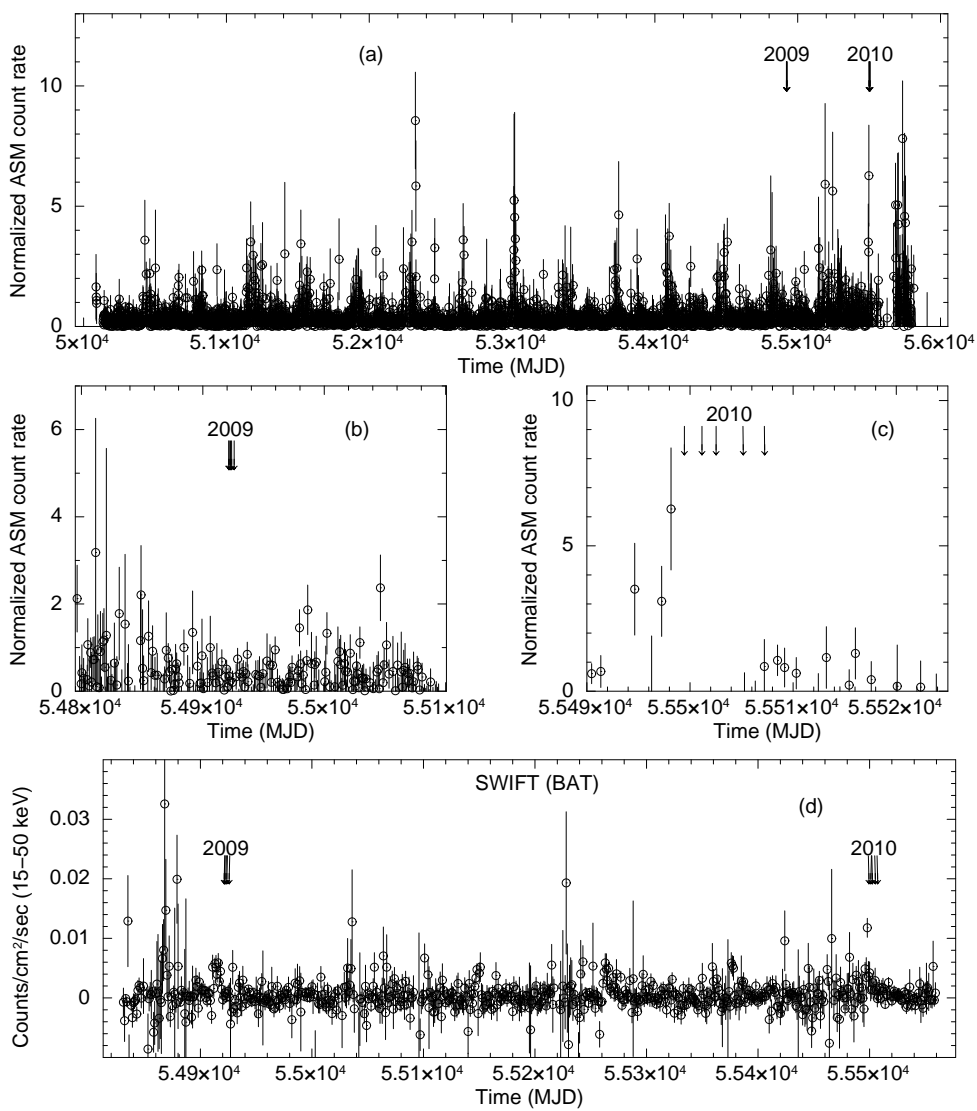

\figurenum{1}
%\epsscale{0.95}
%\vspace{1.2 cm}
%\hspace{0.5 cm}
%\plotone{figure_1_vert.epsi}
\includegraphics[scale=0.5,angle=-90]{fig1a.eps}
\includegraphics[scale=0.5,angle=-90]{fig1b.eps}
%\vspace{-0.75 cm}
\caption{Top panel (a) shows the ASM lightcurve of the source IGR J19294+1816 in 1.5-12 keV energy range. Panel (b) shows the faint long 2009 outburst profile of the ASM lightcurve and panel (c) shows the 2010 outburst of the source.  Bottom panel (d) shows the 2009 and 2010 lightcurve of the source observed in 15-50 keV by SWIFT/BAT hard X-ray transient monitor from MJD 54831-55559. The PCA pointed mode observations are indicated with arrows in all the panels. \label{fig_1}}
\end{figure}

The daily averaged ASM lightcurve of IGR J19294+1816 from MJD 50087 to MJD 55906 in the energy range from 1.5-12 keV is shown in Figure \ref{fig_1}(a). PCA observations during 2009 and 2010 outbursts are indicated by arrows. The lightcurve shows the presence of small flares in every $\sim$ 350 days. Figure \ref{fig_1}(b) presents the 2009 outburst of the source starting from MJD 54796 to MJD 55100 covering 298 days which is the efolding time of the outburst. At the onset of the outburst on MJD 54796 the ASM count rate was 2.12$\pm$0.76 counts/s. PCA observed the source in the decay phase of the 2009 outburst on five occasions from MJD 54921.32 to MJD 54925.82 during which the average PCA flux varied from 6.19$\pm$0.15 counts/s to 3.71$\pm$0.13 counts/s. Similarly Figure \ref{fig_1}(c) shows the ASM lightcurve of the 2010 outburst that started from MJD 55490 (ASM flux = 0.61$\pm$0.34 counts/s). This outburst decayed comparatively faster in $\sim$ 29 days and the ASM count rate declined to 0.14$\pm$0.90 counts/s on MJD 55522. The PCA observations were made on 5 occasions, starting from MJD 55499.46 to MJD 55507.24 during which the average PCA flux varied from 21.22$\pm$0.87 counts/s to 6.80$\pm$0.13 counts/s. Figure \ref{fig_1}(d) shows Swift/BAT hard X-ray transient monitor daily averaged lightcurve \citep{2013ApJS..209...14K} for IGR J19294+1816 in 15-50 keV energy band during MJD 54831-55559.

%PDS flux wise
Lightcurves with 0.01 sec binning was used to generate the power density spectra (PDS) (using ftools task `powspec'), from all the 10 PCA observations. The lightcurves were divided into stretches of 16384 bins per interval. PDS from all the segments were averaged to produce the final PDS for each observation. The PDS of the source exhibits a continuum that is best fit by a power law in the frequency range 5 mHz to 50Hz, and in addition there is a strong peak at $\sim$ $0.0803\pm0.0021$ Hz attributed to the pulsation from the source. The error values are obtained following the standard procedure in XRONOS. A Lorentzian model component is best fit for the observed pulsation peak and its harmonic, whenever present. During the 2009 outburst the pulsations were detected clearly on three occasions and marginally on one occasion on MJD 54921.70, while no pulsation peak was detected at the lowest flux level on MJD 54925.83. During the 2010 outburst prominent pulsation peaks were detected in the PDS in all the 5 observations. Details of the parameters pertaining to the pulsation are presented in Table \ref{fig_1}. 

\begin{figure}
\figurenum{2}
%\epsscale{0.85}
%\vspace{1.2 cm}
%\hspace{0.5 cm}
%\plotone{figure_1_vert.epsi}
\includegraphics[scale=0.4,angle=-90]{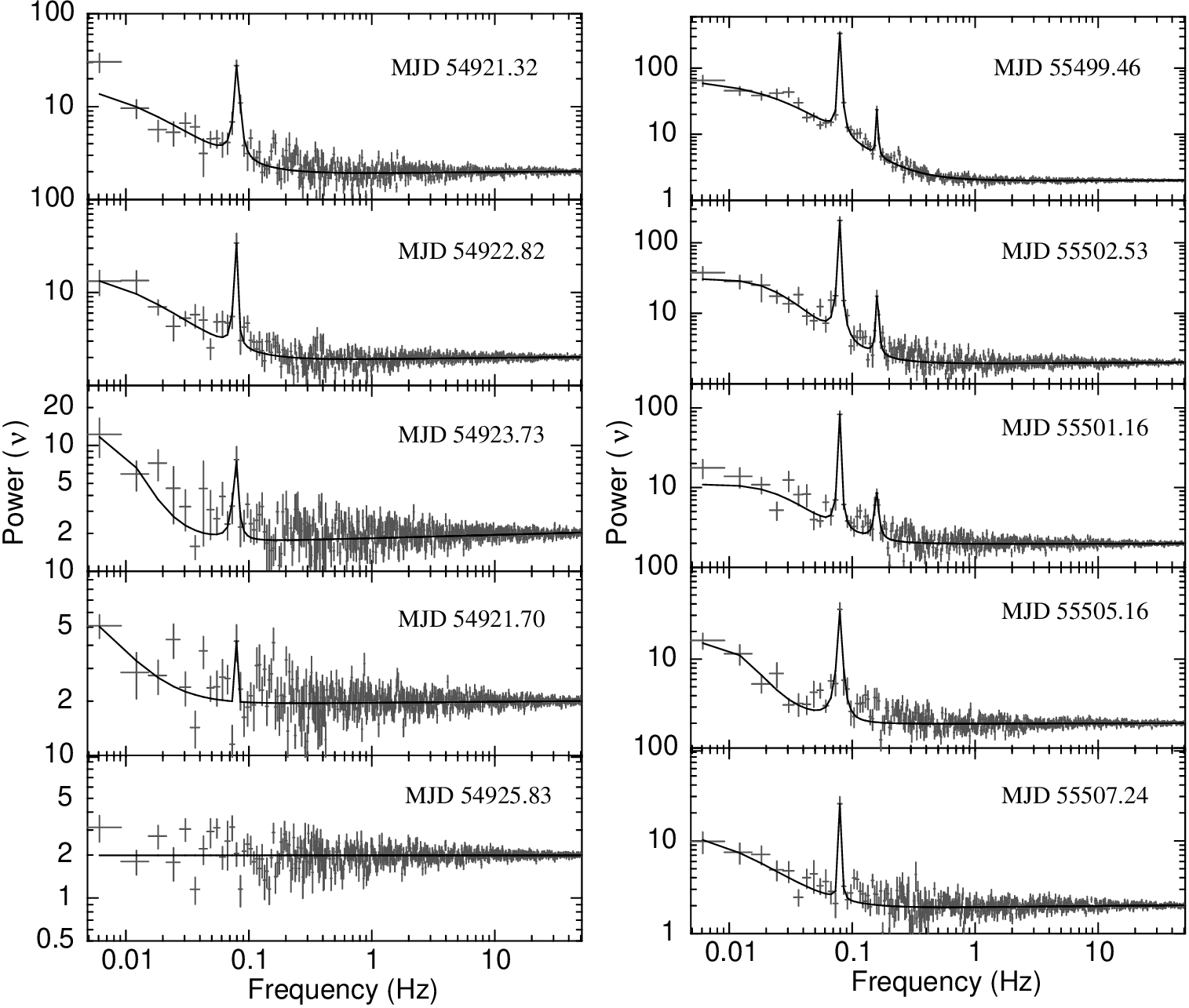}
%\vspace{-0.75 cm}
\caption{The power density spectra (PDS) of the source IGR J19294+1816 from 2009 (left panel a) and 2010 (right panel b) RXTE PCA observations. PDS are arranged with highest to lowest fluxes in both the outbursts. Pulsation peak at $\sim$ $0.0803{\pm0.0021}$ Hz are observed in the PDS except for the lowest flux on MJD 54925.83 during 2009 outburst. Harmonic of the pulse peak $\sim$ $0.160\pm0.004$ Hz along with pulse period are detected from 2010 outburst on MJD 55499.46, 55502.53 \& 55501.16.  \label{fig_2}}
\end{figure}

The PDS obtained from the lightcurves of the 5 observations of 2009 outburst are shown in left column of panels of Figure \ref{fig_2} and those from 2010 outburst are shown in the right column of panels of the same. The PDS corresponding to the highest flux value in the top panel with the other PDS corresponding to progressively lower flux values, as shown in Table \ref{tbl-2} below, are shown in the lower panels. On occasions where the total flux (3-60 keV) is higher then 2.12 $\times$ 10$^{-10}$ ergs cm$^{-2}$ s$^{-1}$ a second peak at $\sim$ $0.160\pm0.004$ Hz, representing the first harmonic of the pulsation peak, is observed.

\begin{figure}
\figurenum{3}
\epsscale{0.85}
%\vspace{1.2 cm}
%\hspace{0.5 cm}
%\plotone{figure_1_vert.epsi}
\includegraphics[scale=0.5,angle=-90]{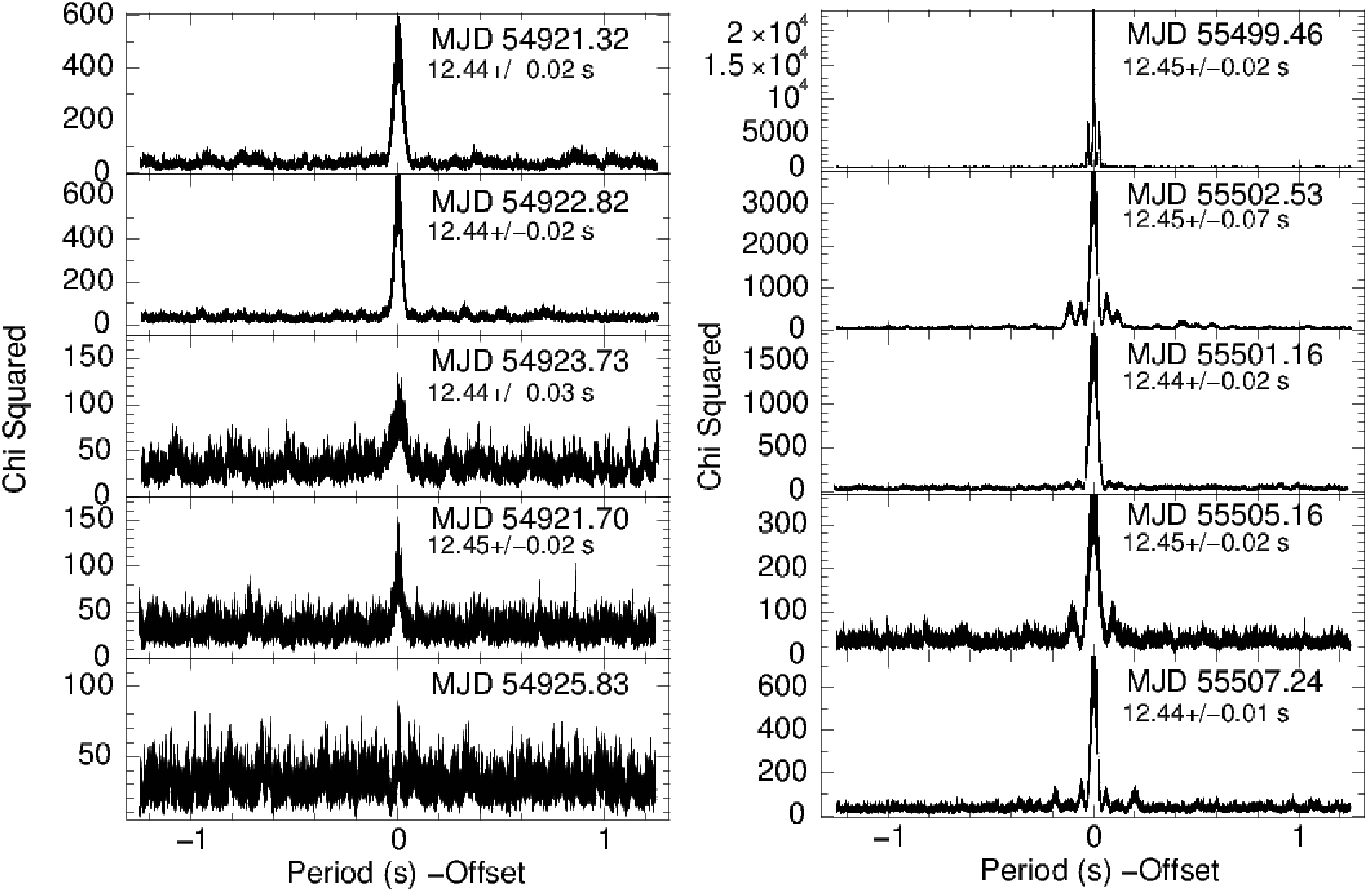}
\caption{The best pulse period of the source IGR J19294+1816 estimated using ftool "efsearch" on 10 PCA observations during 2009 (left panel) \& 2010 (right panel) of the source IGR J19294+1816. Panels are arranged with highest to lowest fluxes in both the outbursts.\label{fig_3}}
\end{figure}

A better estimate of the pulsation period is obtained by $\chi^{2}$ maximization method after folding the lightcurves around an approximate period $\sim$ 12.44 s (as obtained from the PDS) using the ftools task `efsearch'. The lightcurves with a binning of 5 ms were folded with 25000 different periods around 12.44 sec with a resolution of 1 ms with 32 phasebins per period. This pulsation peak was fit with a Gaussian model whose width provided the error on the observed periodicity. The results of this search for pulsation periodicity from all the observations of the two outbursts in 2009 and 2010 are presented in Figure \ref{fig_3}.
A prominent peak corresponding to pulse period $\sim$ 12.44 sec is observed in 2-60 keV energy band of the source except for MJD 54925.83. Best estimated pulse period of all observations are tabulated in Table \ref{tbl-1}.

\begin{figure}
%\vskip{-1.5 cm}
\figurenum{4}
\epsscale{0.80}
%\plotone{figure_2.epsi}
%\vspace{1.2 cm}
%\hspace{0.40 cm}
\includegraphics[scale=0.28,angle=-90]{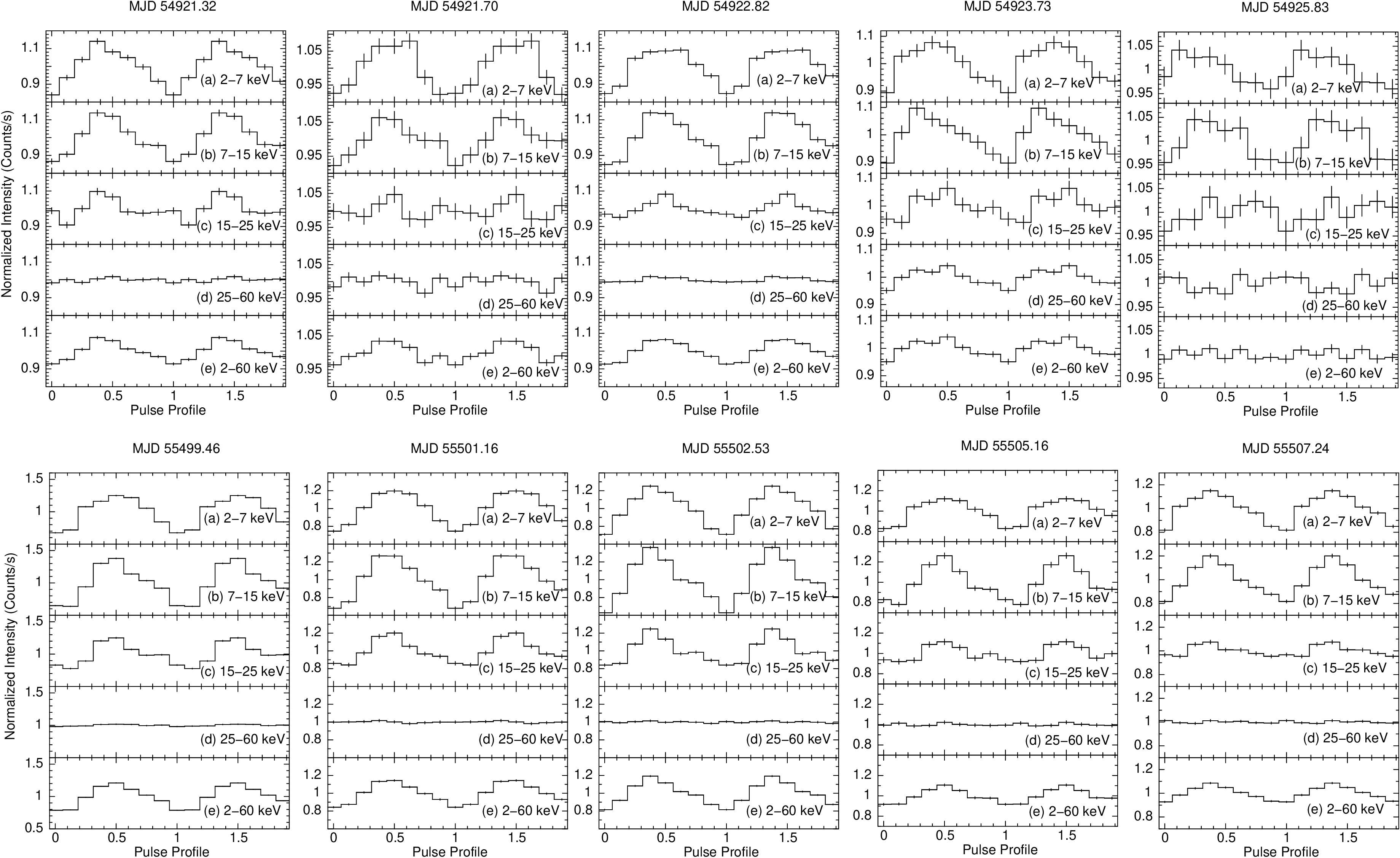}
%\vspace{-0.75 cm}
\caption{The evolution of pulse profile of all the 10 observations RXTE/PCA pointings during the 2009 and 2010 outburst of the source IGR J19294+1816 in panels (a) 2-7 keV,  (b) 7-15 keV, (c) 15-25 keV, (d) 25-60 keV and (e) 2-60 keV energy bands with clear detection of pulsation folded with their estimated pulse periods and with 8 phasebins/period. \label{fig_4}}
\end{figure}

Pulse profiles for each PCA observation were generated by folding the lightcurve using the ftool 'efold' over the exact pulsation period obtained above. The pulse profiles were generated with 8 phase bins per period. Energy dependent pulse profiles were also generated in the range of 2-7 keV, 7-15 keV, 15-25 keV, 25-60 keV and overall 2-60 keV, using the best estimated periods corresponding to different observations. Energy dependent pulse profiles of all the ten PCA observations are shown in Figure \ref{fig_4}. We observed single peaked pulse profiles in 2-7 keV, 7-15 keV, 15-25 keV and 2-60 keV in 9 of the observations (Figure \ref{fig_4}).
While the pulsation was not observed in one case (MJD 54925.83) the pulse profile were obtained by folding the lightcurves at 12.44 sec.

\begin{figure}
%\vskip{-1.5 cm}
\figurenum{5}
%\epsscale{0.80}
%\plotone{figure_2.epsi}
%\vspace{1.2 cm}
%\hspace{0.40 cm}
\includegraphics[scale=0.5,angle=-90]{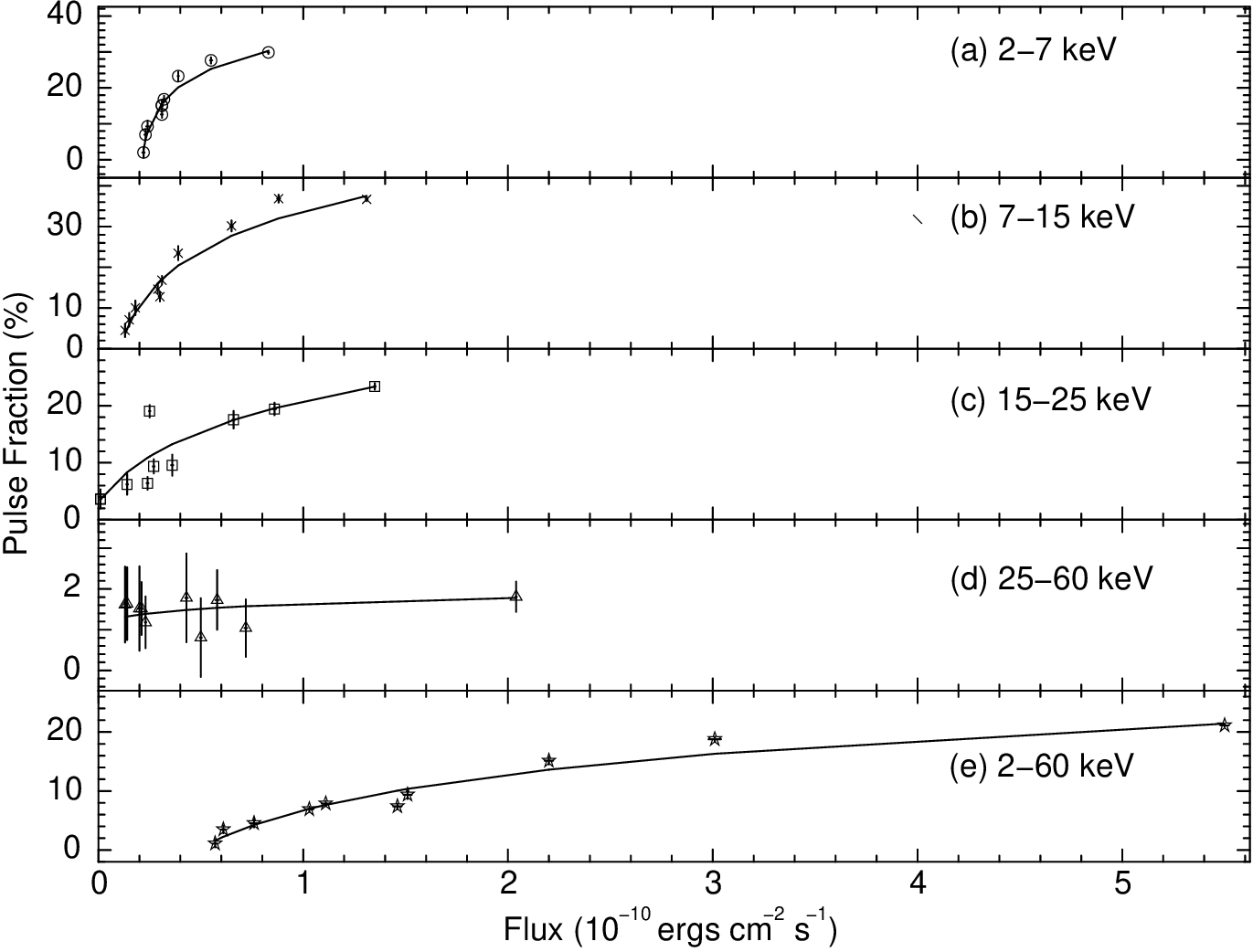}
%\vspace{-0.75 cm}
\caption{The variation of pulse fraction in different energy bands (a) 2-7 keV, (b) 7-15 keV, (c) 15-25 keV, (d) 25-60 keV and (e) 2-60 keV with flux of all the observations (total ten PCA pointings) during 2009 and 2010 outburst of the source IGR J19294+1816. Solid line shows the logarithmic fit to the data. \label{fig_5}}
\end{figure}

This pulse profile was then used to obtain the pulse fraction of the emitted radiation using the traditional method of obtaining the pulse fraction given by  \(\tfrac{C_{max}-C_{min}}{C_{max}+C_{min}}\) \citep{1998ApJ...508..328H}. The evolution of the pulse fraction with the flux of the source in different energy bands, for both the outbursts, is shown in Figure \ref{fig_5}. The pulse fraction shows a logarithmically increasing trend with increasing flux in the energy ranges of 2-7 keV, 7-15 keV, 15-25 keV and 2-60 keV which is a commonly reported pattern for accretion powered pulsars \citep{2004MNRAS.349..173I, 2009MNRAS.395.1662I}. In the 25-60 keV energy band the pulse fraction is very low for all the observations (Figures \ref{fig_4}, \ref{fig_5}). This results in the overall 2-60 keV pulse profile having a comparatively lower pulse fraction (Figure \ref{fig_5}) in all the nine observations as compared to the soft X-ray emissions less then 25 keV. Evidently, from the Figures \ref{fig_2}, \ref{fig_3}, \ref{fig_4} and \ref{fig_5}, the pulsation is not detected for MJD 54925.83 in the total 2-60 keV energy band when the flux was at its lowest. Nevertheless during this observation, the emission in the range of 2-7 keV and 7-15 keV do exhibit a weak single peaked pulse profile when the lightcurve is folded with the average periodicity of 12.44 sec.

\begin{table}
\scriptsize
\begin{center}
\caption{Best period search results of IGR J19294+1816.\label{tbl-1}}
\begin{tabular}{cccclccc}
\tableline
Observation ID & MJD & Date & Exposure &\multicolumn{2}{r}{Power Density Spectra} & Efsearch  &Pulse \\
\cline{5-6}
 & & &&Pulsation&Harmonic &Spin Period &Fraction \\
 & & & (sec)&Peak (Hz)&Peak (Hz)& (Sec)&(\%) \\
\tableline
94103-01-01-00	&54921.32	&2009-03-31~07:42:19.6	&2533&$0.081^{+0.002}_{-0.001}$&---				&12.44$\pm$0.02 &7.38$\pm$0.47\\  
94103-01-01-01	&54921.70	&2009-03-31~16:53:26.6	&3354&$0.079^{+0.058}_{-0.038}$&---				&12.45$\pm$0.02 &3.48$\pm$0.64\\
94103-01-01-03	&54922.82	&2009-04-01~19:37:24.1	&3371&$0.078^{+0.005}_{-0.004}$&---				&12.44$\pm$0.02 &6.91$\pm$0.43\\
94103-01-01-02	&54923.73	&2009-04-02~17:36:24.8	&1448&$0.078^{+0.001}_{-0.001}$&---				&12.44$\pm$0.03 &4.54$\pm$0.71\\
94103-01-02-00	&54925.83	&2009-04-04~19:52:11.4	&3335&---&---							&12.44          &1.01$\pm$0.62\\ 
95438-01-01-00	&55499.46	&2010-10-30~11:03:58.2	&9781&$0.0805^{+0.0004}_{-0.0001}$&$0.160^{+0.002}_{-0.001}$	&12.45$\pm$0.01  &21.12$\pm$0.21	\\ 
95438-01-02-00	&55501.16	&2010-11-01~03:48:31.7	&2935&$0.079^{+0.001}_{-0.001}$&$0.158^{+0.002}_{-0.003}$	&12.44$\pm$0.02 &15.12$\pm$0.60\\
95438-01-02-01	&55502.53	&2010-11-02~12:43:54.3	&2831&$0.079^{+0.001}_{-0.002}$&$0.161^{+0.001}_{-0.001}$	&12.45$\pm$0.07 &18.77$\pm$0.42	\\
95438-01-03-00	&55505.16	&2010-11-05~03:54:04.3	&2328&$0.080^{+0.002}_{-0.001}$&---				&12.45$\pm$0.02 &9.37$\pm$0.72\\
95438-01-03-01	&55507.24	&2010-11-07~05:44:38.1	&3583&$0.080^{+0.001}_{-0.001}$&---				&12.44$\pm$0.01 &7.87$\pm$0.42\\
\tableline
\end{tabular}
\end{center}
\end{table}

\begin{table}
%\scriptsize
\begin{center}
\caption{Observed spectral parameters of IGR J19294+1816.\label{tbl-2}}
%\tiny{
\begin{tabular}{llcccc}
\tableline
& phabs (nH)& Spectral & Flux (3-60 keV)&Powerlaw Flux& Iron line Flux \\
MJD & $10^{22}$ atoms cm$^{-2}$&Index ($\Gamma$)&($10^{-10}erg cm^{-2}s^{-1}$)&($10^{-10}erg cm^{-2}s^{-1}$)&($10^{-12}erg cm^{-2}s^{-1}$) \\
%\cline{2-6} 
%MJD & nH&  $\Gamma$ & Observed & \\ 
\hline
54921.32 &	0.32$^{+1.83}_{-0.32}$	 & $1.51^{+0.14}_{-0.10}$ & 1.48$\pm$0.04& 1.47$\pm$0.04&1.48$\pm$0.57 \\
54921.70 &	0.28$^{+1.23}_{-0.28}$	 & $1.95^{+0.57}_{-1.66}$ & 0.56$\pm$0.02& 0.60$\pm$0.02&1.05$\pm$0.50 \\
54922.82 &	0.97$^{+1.69}_{-0.97}$	 & $1.60^{+0.15}_{-0.13}$ & 1.03$\pm$0.02& 1.37$\pm$0.04&1.63$\pm$0.52 \\
54923.73 &	0.13$^{+3.19}_{-0.13}$	 & $1.78^{+0.27}_{-0.15}$ & 0.72$\pm$0.04& 0.76$\pm$0.04&1.23$\pm$0.71 \\
54925.83 &	0.80$^{+0.79}_{-0.80}$	 & $2.07^{+0.14}_{-0.12}$ & 0.51$\pm$0.02& 0.51$\pm$0.02&1.03$\pm$0.48 \\

55499.46 &	2.9$^{+0.4}_{-0.6}$	 & $1.23^{+0.03}_{-0.05}$ & 5.39$\pm$0.03& 8.05$\pm$0.05&2.17$\pm$0.44 \\
55501.16 &	4.1$^{+1.8}_{-1.8}$	 & $1.22^{+0.11}_{-0.12}$ & 2.12$\pm$0.03& 4.05$\pm$0.06&1.51$\pm$0.55 \\
55502.53 &	3.2$^{+1.5}_{-1.3}$	 & $1.19^{+0.13}_{-0.10}$ & 2.65$\pm$0.03& 5.63$\pm$0.08&2.13$\pm$0.74 \\
55505.16 &	1.7$^{+2.0}_{-1.6}$	 & $1.35^{+0.16}_{-0.21}$ & 1.44$\pm$0.04& 2.21$\pm$0.06&1.51$\pm$0.64 \\
55507.24 &	2.8$^{+1.1}_{-1.4}$	 & $1.76^{+0.07}_{-0.10}$ & 1.14$\pm$0.03& 1.39$\pm$0.03&0.77$\pm$0.52 \\
\tableline 
\end{tabular}
%\vspace{-0.1 cm}
\end{center}
\end{table}

\begin{table}
\small
\begin{center}
\caption{Observed best fit parameters all the 10 observations of IGR J19294+1816 for Cyclotron line and Iron line .\label{tbl-3}}
%\tiny{
\begin{tabular}{ccccccccc}
\tableline
MJD& \multicolumn{2}{c}{Cyclabs}& \multicolumn{3}{c}{Iron line}&    $\chi^{2}$ (dof) & $\chi^{2}$ (dof)& $\chi^{2}$ (dof) \\
& Energy  &Depth&Width&Energy&norm& & without & without \\
\cline{2-4}
\cline{5-6}
 & E$_{cycl}$ (keV)&D$_f$&W$_f$ (keV)&E$_{Fe}$ (keV) &($\times$10$^{-4}$)   &&cyclabs&Fe line  \\ 
\hline
54921.32 &	35.5	& $0.01^{+3.11}_{-0.01}$ 	&5.45   	  &$6.40^{+0.27}_{-0.25}$&$1.4^{+0.7}_{-0.7}$ &51.7 (84)&51.7 (85)&64.7 (86) \\
54921.70 &	35.5	& $0.1^{+5.0}_{-0.1}$		&5.45		  &$6.55^{+0.30}_{-0.30}$&$1.0^{+0.5}_{-0.6}$ &68.6 (82)&68.4 (85) &79.6 (84) \\
54922.82 &	35.5	& $2^{+5}_{-2}$ 		&5.45	 	  &$6.21^{+0.28}_{-0.24}$&$1.7^{+0.6}_{-0.7}$ &68.5 (84) &71.3 (85) &86.4 (86) \\
54923.73 &	35.5	& $0.5^{+9.1}_{-0.5}$ 		&5.45		  &$6.50^{+0.27}_{-0.28}$&$1.2^{+0.8}_{-0.8}$ &59.7 (85) &59.7 (85) &65.8 (86) \\
54925.83 &	35.5	& ---			        &5.45 	          &$6.65^{+0.12}_{-0.17}$&$0.96^{+0.29}_{-0.30}$&62.2 (82) &62.3 (85) &71.8 (86) \\

\textbf{55499.46} &	35.5$^{+2.1}_{-1.7}$	 & $2.10^{+2.0}_{-0.8}$   &$5.45^{+3.10}_{-1.98}$&$6.40^{+0.10}_{-0.15}$&$2.1^{+0.6}_{-0.4}$ &77.9 (82) &\textbf{216.0 (85)} &122.6(84)  \\
55501.16 &	38$^{+22}_{-4}$ & $3.5^{+2.1}_{-1.6}$	&5.45 		  &$6.38^{+0.21}_{-0.24}$&$1.5^{+0.7}_{-0.8}$ &82.6 (82) &114.8 (84) &94.2 (84)  \\
55502.53 &	41$^{+20}_{-9}$ & $4.2^{+2.6}_{-2.0}$	&5.45 		  &$6.31^{+0.24}_{-0.24}$&$2.1^{+0.9}_{-1}$ &68.1 (84)&102.7 (86) &81.4 (86)  \\
55505.16 &	35$^{+14}_{-6}$ & $2.9^{+5.5}_{-2.6}$	&5.45	 	  &$6.15^{+0.50}_{-0.41}$&$1.5^{+0.8}_{-0.8}$ &58.4 (84)&62.3 (86) &67.7 (86) \\
55507.24 &	35.5    	& $1.1^{+5.0}_{-1.1}$	&5.45	 	  &$6.62^{+0.23}_{-0.29}$&$0.73^{+0.35}_{-0.38}$ &59.8 (83)&60.5 (84) &63.7 (85)  \\
\tableline 

\end{tabular}
\flushleft{All values without errors are freezed values. All uncertainties are expressed as 90\% confidence.}
\end{center}
\end{table}

\subsection{Spectral properties}
\begin{figure}
%\vskip{-1.5 cm}
\figurenum{6}
%\vspace{2.30 cm}
\includegraphics[scale=0.65,angle=-90]{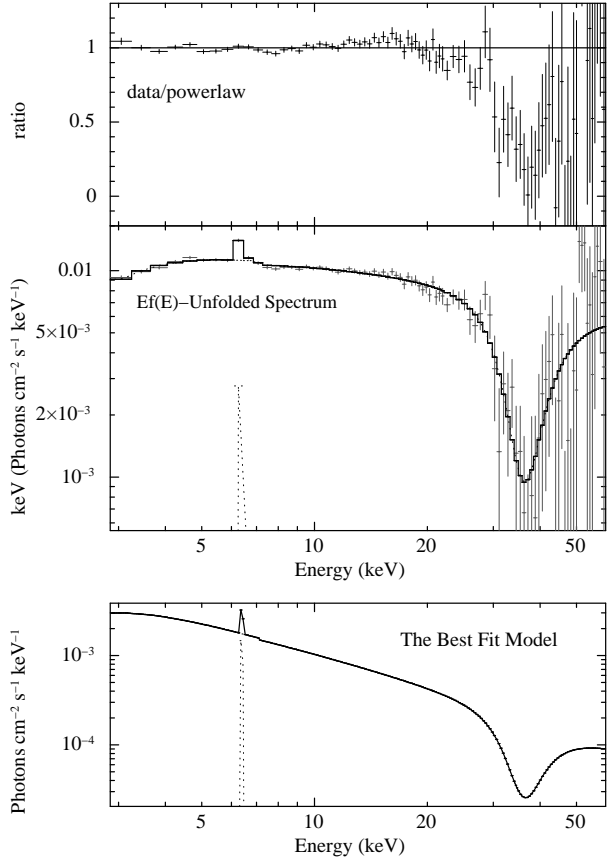}
%\vspace{-1.8 cm}
\caption{The spectra of the observation on MJD 55499.46, Obs. Id.: 95438-01-01-00. The absorption at 35.5$^{+2.1}_{-1.7}$ keV and the Fe line at 6.40$^{+0.10}_{-0.15}$ keV are the two prominent features of the observation on this particular day.\label{fig_6}}
\end{figure}

\begin{figure}
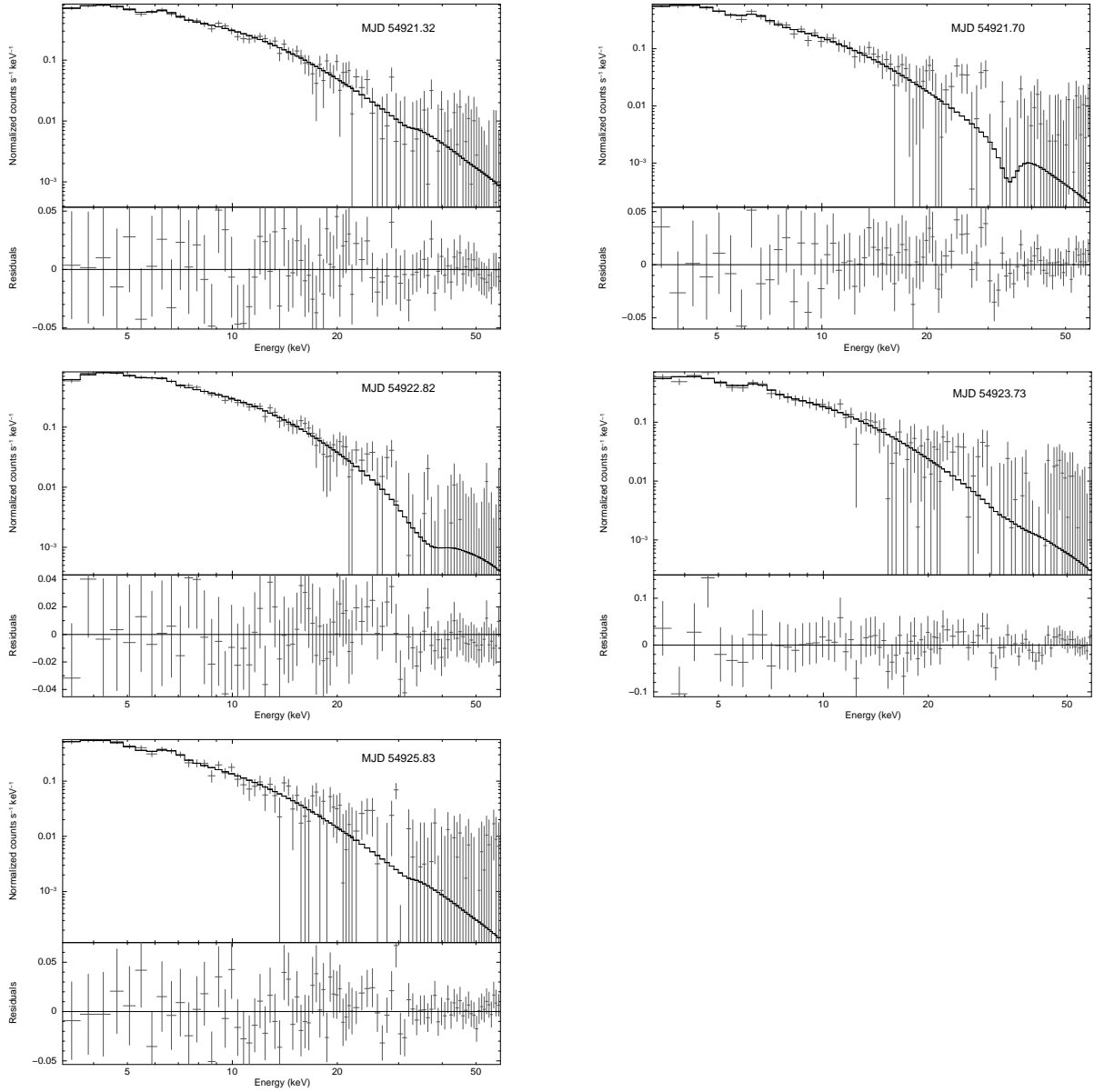

%\vskip{-1.5 cm}
\figurenum{7}
%\vspace{2.30 cm}
\includegraphics[scale=0.3,angle=-90]{fig7a.eps}
\includegraphics[scale=0.3,angle=-90]{fig7b.eps}
\includegraphics[scale=0.3,angle=-90]{fig7c.eps}
\includegraphics[scale=0.3,angle=-90]{fig7d.eps}
\includegraphics[scale=0.3,angle=-90]{fig7e.eps}
%\vspace{-1.8 cm}
\caption{Energy spectrum of IGR J19294+1816 obtained with the 5 PCA/RXTE observations (MJD 54921.32, 54921.70, 54922.82, 54923.73 \& 54925.83) during 2009 outburst of the source along with the best fit model phabs*(powerlaw + gaussian)*cyclabs indicated by solid line. Bottom panel shows the residuals to the best fit model for each observation.\label{fig_7}}
\end{figure}

\begin{figure}
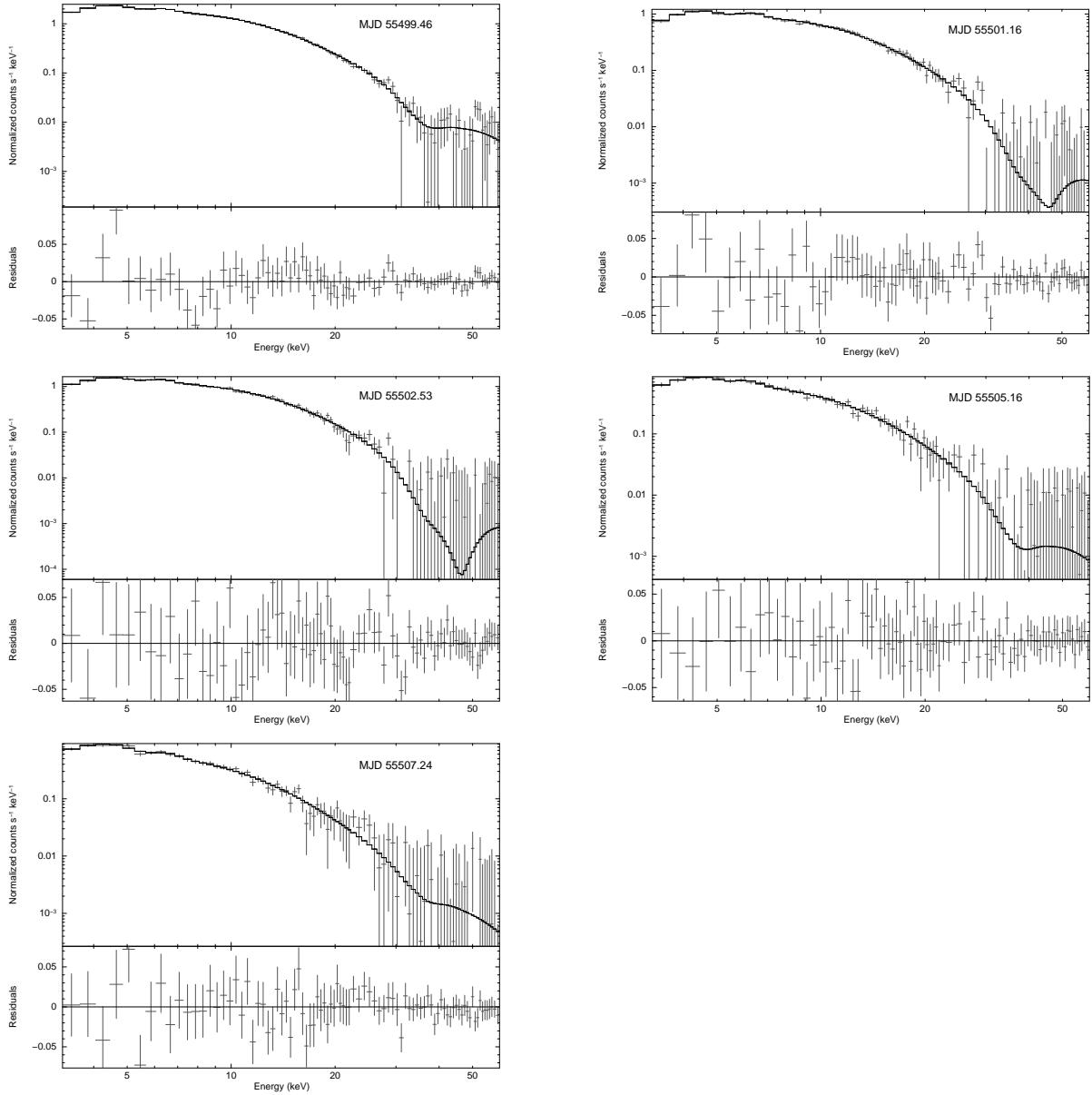

%\vskip{-1.5 cm}
\figurenum{8}
%\vspace{2.30 cm}
\includegraphics[scale=0.3,angle=-90]{fig8a.eps}
\includegraphics[scale=0.3,angle=-90]{fig8b.eps}
\includegraphics[scale=0.3,angle=-90]{fig8c.eps}
\includegraphics[scale=0.3,angle=-90]{fig8d.eps}
\includegraphics[scale=0.3,angle=-90]{fig8e.eps}
%\vspace{-1.8 cm}
\caption{Energy spectrum of IGR J19294+1816 obtained with the 5 PCA/RXTE observations (MJD 55499.46, 55501.16, 55502.53, 55505.16 \& 55507.24) during 2010 outburst of the source along with the best fit model phabs*(powerlaw + gaussian)*cyclabs indicated by solid line. Bottom panel shows the residuals to the best fit model for each observation.\label{fig_8}}
\end{figure}

The 3-60 keV wide band continuum X-ray spectrum of the source is fit by a simple `power law' model (Table \ref{tbl-2}), with an absorption component for the Galactic inter stellar medium parameterized by the spectral  model component `phabs' (available in XSPEC package) which provides the measure of absorption as
the effective equivalent hydrogen column density (in units of 10$^{22}$ atoms/cm$^{2}$) using the photoelectric absorption cross section of \citet{1992ApJ...400..699B}. In addition we used a Gaussian line to account for the statistically significant presence of the 6.4 keV iron fluorecence line produced in the accretion process. The width of the iron line is fixed at 0.01 keV. On MJD 55499.46, corresponding to the maximum value of the measured flux the fit to the simple model yielded a high chi-square and the residuals indicated a possible absorption feature at about $\sim$ 35 keV (MJD 55499.46 is indicated in bold fonts in Table \ref{tbl-3}). The spectrum of this particular observation is shown in Figure \ref{fig_6}. The top panel shows the ratio of the spectrum to the best fit powerlaw model, and the residual absorption feature is best modeled by `cyclabs' model component which signifies the physical presence of cyclotron resonance scattering features (CRSF). The middle panel of Figure \ref{fig_6} depicts the \texttt{XSPEC}'s unfolded spectra. The bottom panel shows the model that best fits the spectrum of this observation. For consistency we have used the same model (phabs*(powerlaw + gaussian)*cyclabs), in the XSPEC terminology, to fit all the ten observations during 2009 and 2010 outbursts i.e, including the iron line and cyclotron line. The folded energy spectrum of IGR J19294+1816 obtained with the five PCA/RXTE observations each during 2009 and 2010 outburst of the source along with the best fit model indicated by solid line are shown in Figure \ref{fig_7} and \ref{fig_8} respectively. Bottom panel of each of the spectrum of Figure \ref{fig_7} and \ref{fig_8} gives the residuals to the best fit model for each observed spectra.

To estimate the flux and error on flux in the 3-60 keV range, an additional  model component `cflux' (convolution model in XSPEC) is used to obtain the overall flux as well as the flux pertaining to the individual modeled component, viz unabsorbed powerlaw, Gaussian line, etc. The best fit values of the model and the flux obtained from the spectra are tabulated in Tables \ref{tbl-2} and \ref{tbl-3}.  In Table \ref{tbl-3} the value of $\chi^{2}$ and degree of freedom without the inclusion of `cyclabs' model used for modeling the CRSF is also reported. We observed that the inclusion of cyclotron line model in the spectra of the three bright observations (MJD 55499.46, 55501.16 and 55502.53) during 2010 outburst showed significant improvement in the $\chi^{2}$ and degree of freedom. Similarly, the value of $\chi^{2}$ and degree of freedom without the inclusion of Gaussian model is also reported in the Table \ref{tbl-3}. This provides an estimate of the significance of the respective model components for each spectra.

\begin{figure}
%\vskip{-1.5 cm}
\figurenum{9}
%\vspace{2.30 cm}
\includegraphics[scale=0.5,angle=-90]{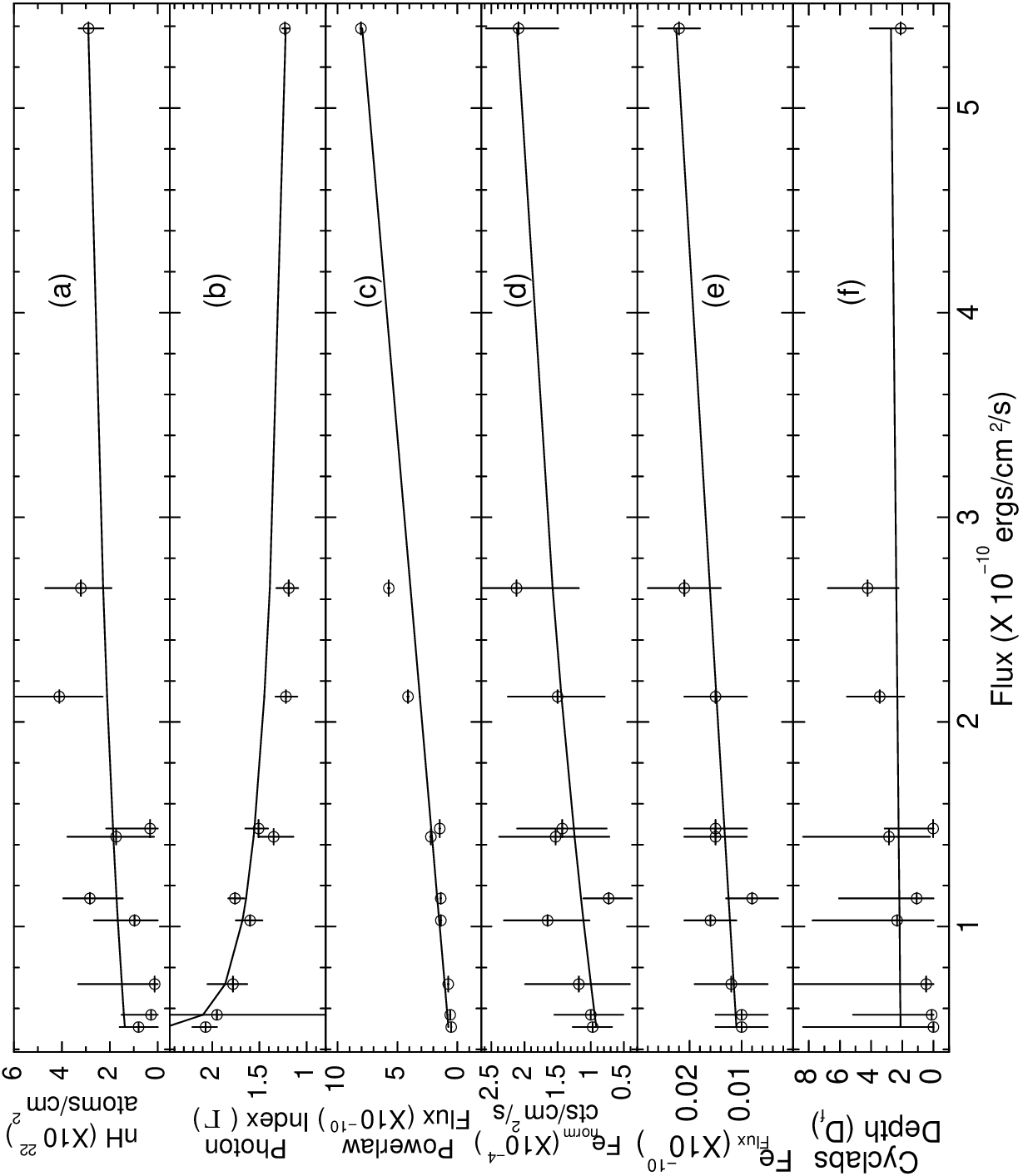}
%\vspace{-1.8 cm}
\caption{The variation of spectral parameters, (a) hydrogen column density (in units of $10^{22}$ atoms cm$^{-2}$), (b) Photon Index ($\Gamma$), (c) Powerlaw flux (in the units of ergs/cm$^2$/s), (d) Iron line normalization, (e) Iron line flux (in the units of ergs/cm$^2$/s), (f) Cyclotron line depth with 3-60 keV absorbed total flux from 10 observations. Solid lines indicate the fit of linear trend in panels (a), (c), (d) and (e). Solid line in Panel (b) shows logarithmic trend used to fit the spectral correlation. \label{fig_9}}
\end{figure} 

The variation of different spectral parameters with total flux in 3-60 keV energy range is shown in Figure \ref{fig_9}. We obtained Pearson's correlation coefficient to quantify the strength of the relation of spectral parameters with respect to total flux. There seem to be an increase in the absorption in the source, as parameterized by the nH value (Figure \ref{fig_9}a), with increase in flux, with a correlation coefficient of 0.59 (the obvious assumption is that the properties of ISM does not change in tandem with the source). Figure \ref{fig_9}(b) shows that the photon index ($\Gamma$) shows spectral hardening with increasing 3-60 keV flux with a relatively strong anti-correlation as the Pearson's coefficients has a value of -0.72. A logarithmic function is used to fit the trend of spectral index variation with the flux. This anti-correlation suggests that the source is in horizontal branch of the hardness intensity diagram \citep{reig2013}, further strengthening the hypothesis that the nature of the binary system is that of the Be/X-ray binary type. The flux corresponding to the spectral component `powerlaw' (Figure \ref{fig_9}c) is mostly the unabsorbed flux for this source which is consistently higher by about 0.51$\pm$0.02$\times$$10^{-10} ergs/cm^2/s$ in nearly all observations and linearly increasing with increase in total flux of the source with a strong positive correlation of 0.96.
Figure \ref{fig_9}(d) shows a trend of linear increase of the iron line normalization with the total flux, the correlation coefficient is 0.76. The iron line flux shows a positive correlation coefficient of 0.82 (Figure \ref{fig_9}e) with increasing total flux. Reprocessing of the hard X-ray continuum in relatively cool matter that produces the fluorescent iron K$\alpha$ line feature does not vary with flux, as observed from Table \ref{tbl-3}. The CRSF parameters were not significant enough in all the observations and hence a correlation test was not possible for the most important spectral feature detected.

\section{Results \& Conclusion}

In this work we have studied the timing and spectral properties of Be/X-ray binary IGR J19294+1816 using RXTE/PCA data. The RXTE/PCA observed the evolution of the source in the decaying phase for both the outbursts in 2009 and 2010. The overall spectral and timing features are very similar during these two outbursts, suggesting that the physical processes in the accretion region during the two outbursts separated by $\sim 600$ days are similar. We observed that the significance of the detection of the pulsation peak decreases with the decrease in flux. Furthermore, the pulse fraction showed logarithmically increasing trend with flux in all the energy ranges (as seen in Figure \ref{fig_5}). 

The spectral study reveals a detection of a CRSF for the first time in the source IGR J19294+1816 at 35.5$^{+2.1}_{-1.7}$ keV. In addition, an iron Fe line at 6.40$^{+0.10}_{-0.15}$ is also reported to be present in the source. Cyclotron absorption features originates in the X-ray spectrum due to the resonance scattering of photons with quantized electrons in the presence of magnetic field. The presence of the cyclotron absorption lines enable the direct measurement of the magnetic field of the pulsar as given by the following relation \citep[see page 471][]{Accretion_1}:
\begin{equation}
E_c = 11.6B_{12}(1+z)^{-1} keV,
\label{eq1}
\end{equation}
where $E_c$ is the energy of the cyclotron absorption line, $z$ is the gravitational red shift at the neutron star surface and $B_{12}$ is the magnetic field in units of $10^{12}$ Gauss. Typically, the value of $z$ is 0.35 for neutron stars of mass in the range $1.4-2 M_{\odot}$ \citep{2002Natur.420...51C, 2014RMxAA..50..103Z}. Hence, using the value of $E_c = 35.5$ keV (corresponding to the most significant observation of the CRSF), the value of magnetic field obtained is $B = 4.13\times10^{12}$ Gauss.

The detection of the CRSF at lower flux values is marginal at best fit, as a result there is no clear correlation between the energy of the cyclotron line and the X-ray luminosity (Table \ref{tbl-3}), similar to sources like 1A 0535+262 \citep{2007ESASP.622..471C}. Although there is mild positive correlation (coefficient of 0.41) between the cyclotron line depth and the total flux (it may be noted that in most of the cases energy of the CRSF is freezed see Table \ref{tbl-3}), it is not much of consequence as the error on the best fit value shows that the statistically significant detection of the CRSF occurs only at the highest flux value. Since the powerlaw hardens as the flux increases, one factor for significant detection of the CRSF at high flux is that the comparatively harder spectra provides the better continuum baseline above the background noise to enable a statistically significant detection of the CRSF.

%%%%%%%%%%%%%%%%%%%%%%%%%%%%%%%%%%%%%%%%%%%%%%

\section{Acknowledgment}
This research has made use of data obtained through the HEASARC Online Service, provided by the NASA/GSFC, in support of NASA High Energy Astrophysics Programs. JR and PCA acknowledge the fellowship and the funding provided by the National Academy of Sciences, India (NASI). MC and JR acknowledges the many discussion with suggestions provided by Professor A. R. Rao, TIFR, Mumbai, India. We also thank the referee for his detailed valuable suggestions which considerably improved the manuscript.
\bibliographystyle{apj}
%\bibliography{bibtex.bib}

\end{document}